\documentclass[%
 reprint,
 amsmath,amssymb,
 aps,
 prl,
]{revtex4-2}

\usepackage{physics}
\usepackage{graphicx}
\usepackage{dcolumn}
\usepackage{bm}
\usepackage{siunitx}

\usepackage[whole]{bxcjkjatype}
\usepackage{color}

\usepackage{mathrsfs}

\usepackage{txfonts}

\begin{document}

\preprint{APS/123-QED}

\title{Photocurrent induced by a bicircular light drive in centrosymmetric systems
}%

\author{Yuya Ikeda}
\affiliation{%
 Department of Applied Physics, The University of Tokyo, Hongo, Tokyo, 113-8656, Japan
}%

\author{Sota Kitamura}
\affiliation{%
 Department of Applied Physics, The University of Tokyo, Hongo, Tokyo, 113-8656, Japan
}%

\author{Takahiro Morimoto}
\affiliation{%
 Department of Applied Physics, The University of Tokyo, Hongo, Tokyo, 113-8656, Japan
}%

\date{\today}

\begin{abstract}
A bicircular light (BCL) consists of left and right circularly polarized lights with different frequencies, and draws a rose-like pattern with a rotational symmetry determined by the ratio of the two frequencies.  
Here we show that an application of a BCL to centrosymmetric systems allows a photocurrent generation through introduction of an effective polarity to the system.
We derive formulas for the BCL-induced photocurrent from a standard perturbation theory, which is then applied to a simple 1D model and 3D Dirac/Weyl semimetals. 
A nonperturbative effect with strong light intensity is also discussed with the Floquet technique.
\end{abstract}

\maketitle





\begin{figure}[t]
\begin{center}
\includegraphics[height=6cm]{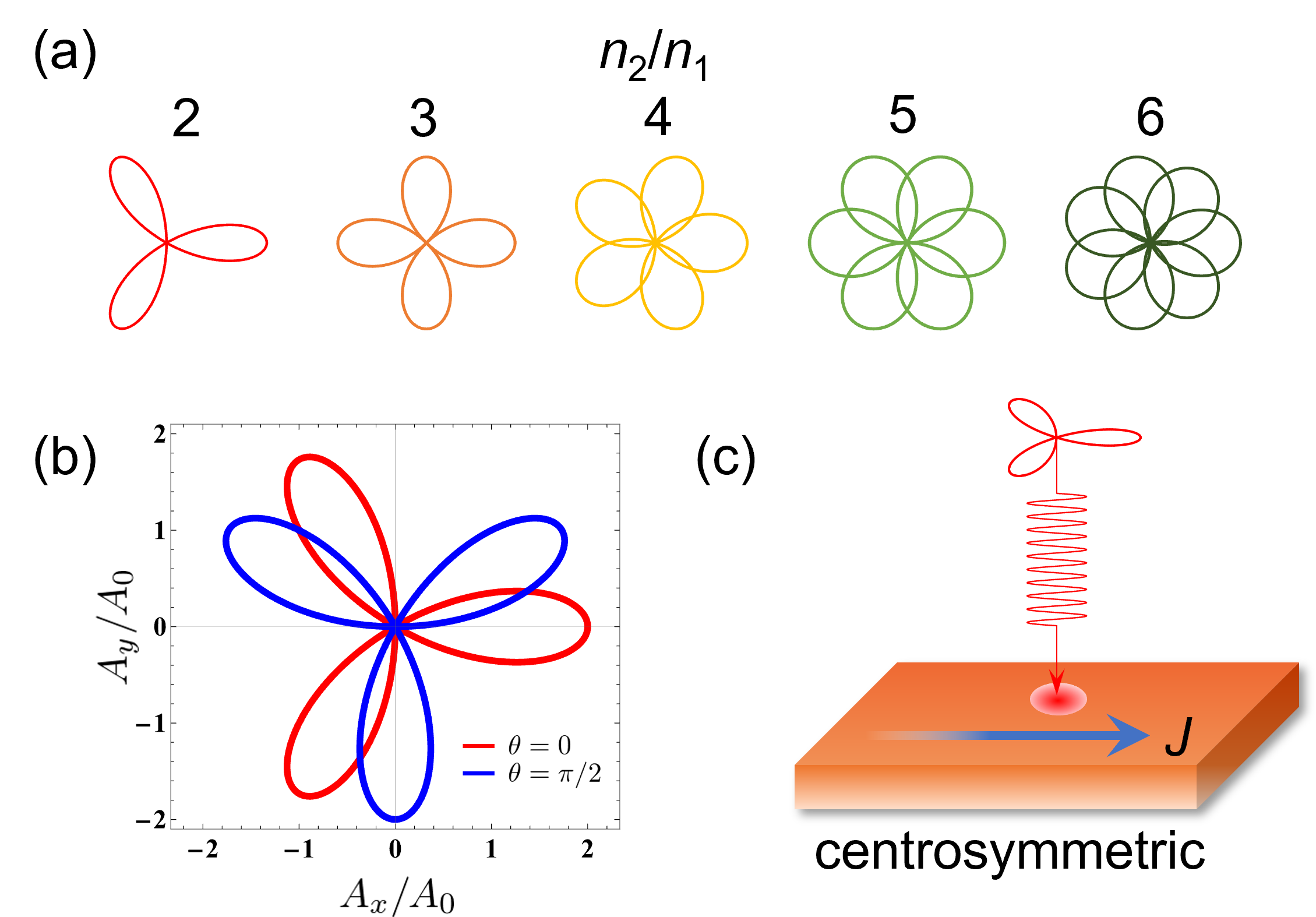}
\caption{Schematics of bicircular light (BCL) that consists of circularly polarized light (CPL) with different frequencies.
(a) The electric field of the BCL draws the rose-like pattern with a rotational symmetry determined by the ratio of the two frequencies. 
The rose patterns with $n_1=1, n_2=n$ in Eq.~\eqref{eq: rose pattern} are plotted. 
(b) The relative phase between two CPL waves leads to rotation of the rose pattern of the BCL.
(c) Application of BCL to centrosymmetric systems can induce photocurrent with introduction of an effective polarity.
}
\label{fig:1}
\end{center}
\end{figure}

\textit{Introduction. ---}
Symmetry plays a central role in studying quantum phases of matter and governs their responses to external perturbations \cite{ryu-rmp}. For example, time-reversal symmetry breaking allows quantum Hall effects with quantized Hall conductivity \cite{nagaosa-rmp}. Inversion symmetry breaking is necessary for emergence of electric polarization \cite{resta-RMP} and bulk photovoltaic effect \cite{Sturman}. 
Symmetry of the electronic system in solids is usually determined by the crystal structure, spontaneous symmetry breaking in the ground state such as magnetic order, and an application of the external field.

Dynamical control of quantum systems by periodic driving has attracted keen attention and is recently called ``Floquet engineering'' \cite{Oka-Kitamura19,Rudner2020,andre-RMP}. 
Periodic driving has an advantage that it can control the symmetry and topology of quantum materials without changing their chemical composition and sometimes offers novel quantum phases which have no counterpart in the equilibrium \cite{Lindner2011,Rudner2013,Zhang2016,Hu2020,Kitamura2022}.
In particular, an application of the circularly polarized light (CPL) can introduce an effective time-reversal symmetry (TRS) breaking to the system, which is exemplified by the emergence of an anomalous Hall insulating phase in graphene under CPL \cite{Oka2009,Wang2013}.


Employing a two-frequency drive instead of a monochromatic drive has recently been attracting an interest as a method to enlarge the capabilities of Floquet engineering \cite{Neufeld2019,Kang20,Gopal2021,Sandholzer22,Minguzzi22,Gopal2022,Wang2023}. In particular, the spatial inversion symmetry and rotational symmetry of the system can be controlled by applying so called bicircular light (BCL). The BCL consists of two CPL waves with different frequencies and opposite chirality \cite{Kfir2015}, and is expressed in the form of a vector potential 
as
\begin{align} 
    A_x(t)+ i A_y(t) = A_0 [e^{in_1\Omega t} +  e^{-in_2\Omega t+i\theta}],
    \label{eq: rose pattern}
\end{align}
where $A_0$ is the amplitude and $n_1,n_2$ are the integers representing the frequencies of the two CPL waves. The BCL wave draws the rose-like pattern with $(n_1+n_2)/\textrm{gcd}(n_1,n_2)$-fold rotational symmetry as shown in Fig.~\ref{fig:1}(a), which implies that applying BCL can reduce the system's rotational symmetry from that of the crystal structure. The parameter $\theta$ is the phase difference between two CPL waves and serves as a knob to rotate the rose pattern drawn by the BCL wave [see Fig.~\ref{fig:1}(b)].
Recently, several studies on the control of symmetry and topology using a BCL drive have been reported, including charge dynamics in graphene \cite{Nag19} and Weyl point generation in Dirac semimetals \cite{Trevisan22}.
Moreover, it was also shown that the BCL driving can introduce polarity in centrosymmetric systems and induce electric polarization due to inversion symmetry breaking originating from incompatible rotation symmetry of the BCL~\cite{Ikeda2022}.

BCL's ability to control the inversion and rotation symmetry of the materials is also expected to apply to optoelectronic properties of materials. 
In particular, bulk photovoltaic effect (BPVE) \cite{Sturman} is an important nonlinear optical phenomenon in a system lacking inversion symmetry, where the light irradiation induces a dc photocurrent \cite{Grinberg2013,Nie2015,tan2016shift}.
There are several mechanisms for BPVE, including the shift current, injection current, and ballistic current \cite{dai2022recent, Nagaosa-Morimoto17, Orenstein21}. Among them, the injection current is a photocurrent proportional to relaxation time of photocarriers and plays a dominant role in the circular photogalvanic effect in which circularly polarized light induces chirality dependent photocurrent \cite{Orenstein21}.
Indeed, application of two frequency drive has been studied as a method to induce injection current \cite{Sipe1996} and injection spin current \cite{Sipe2000} based on a perturbation theory, showing a coherent control of the current direction with the relative phase of the two frequency lights.
Such coherent control of photocurrent was demonstrated in Bi$_2$Se$_3$ in a collinear polarization scheme \cite{Bas2015}.
More recently, photocurrent induced by co-rotating CPL in the strong intensity regime has been studied with an ab initio calculation \cite{Neufeld2021}.

In this paper, we study photocurrent induced by the BCL, on the basis of Floquet engineering of inversion and rotational symmetries of the system.
We demonstrate that irradiating the $C_3$-symmetric BCL creates photocurrent in inversion symmetric systems such as $C_2$ or $C_4$ symmetric systems [Fig.~\ref{fig:1}(c)],
where photocurrent cannot be induced by conventional monochromatic light irradiation. Specifically, using the perturbation theory and the Floquet theory, we derive a formula for the BCL-induced photocurrent which is proportional to the relaxation time $\tau$ (i.e., injection current) in systems with spatial symmetries. We apply the obtained formula to a 1D system with inversion symmetry and a 3D Dirac/Weyl semimetal, where the latter exhibits a large photocurrent due to its gapless nature. The direction of the photocurrent can be controlled by the pattern of the BCL light through the phase $\theta$. Also, nonperturbative effects on the photocurrent with the light intensity are discussed with the Floquet-Keldysh formalism.

\textit{BCL-induced photocurrent. ---}
Let us study photocurrent induced by a BCL drive in centrosymmetric systems based on a standard perturbation theory.
Specifically, we adopt a diagrammatic approach \cite{Parker19}
to calculate the nonlinear optical conductivity for the BCL-induced photocurrent. 
Under the vector potential $\bm{A}(t)$ of the BCL, the dynamics of the electron obeys the time-dependent Schr\"{o}dinger equation with the time-dependent Hamiltonian $H(t)$ with the minimal coupling, 
\begin{align}
    H(t)=H_0(\vb*{k}+e\vb*{A}(t)/\hbar),
\end{align}
where $H_0(\vb*{k})$ is the Bloch Hamiltonian in the equilibrium.
In the following, we focus on the $C_3$-symmetric BCL represented by $\vb*{A}(t) \equiv (A_x(t),A_y(t))$ with $(n_1,n_2)=(1,2)$ in Eq.~\eqref{eq: rose pattern}.
The electric field is given by
$\vb*{E}(t)=-\partial\vb*{A}(t)/\partial t\equiv 
\mathrm{Re}[\vb*{E}^{(\Omega)} e^{i\Omega t}+\vb*{E}^{(-2\Omega)} e^{-2i\Omega t} ]
$,
with the complex electric fields $\vb*{E}^{(\Omega)}$ and $\vb*{E}^{(-2\Omega)}$.
BCL driving breaks both time-reversal symmetry and spatial symmetries including inversion,
which leads to a photocurrent generation even in systems with spatial inversion or rotational symmetry.
Specifically, the BCL-induced photocurrent
is a third-order response with respect to the electric field at the lowest order as 
with the combination of the photon energies, $\Omega+\Omega-2\Omega=0$.
The dominant contribution to the photocurrent is the so called injection current contribution which is proportional to the relaxation time $\tau\sim \hbar/\gamma_0$.
In the diagrammatic approach, such contribution is included in the box diagram as detailed in Appendix. 
By keeping only the terms $\propto 1/\gamma_0$ in the diagrammatic computation,  
we obtain the expression for the BCL-induced photocurrent as 
\begin{align}
    J^{\mu\alpha\beta\gamma}_{\rm BCL}(\Omega)=\Re \left[
    \sigma^{\mu\alpha\beta\gamma}_{\rm BCL}
     E^{(\Omega)}_\alpha E^{(\Omega)}_\beta E^{(-2\Omega)}_\gamma \right]
\end{align}
with
\begin{align}
    \sigma_{\rm BCL}^{\mu\alpha\beta \gamma}
    &\simeq \frac{i\pi e^4}{3\hbar^4 \Omega^3\gamma_0}
    \sum_{\substack{\{\alpha,\beta,\gamma\}\\ a,b,c}}\int [\dd \vb*{k}] f_{ab}(v_{aa}^\mu-v_{bb}^\mu) 
    \nonumber \\
    &\times\Bigg[
     \left( \frac{v_{ac}^\alpha v_{cb}^\beta v_{ba}^\gamma}{\epsilon_{ac}+2\hbar\Omega}+\frac{v_{ac}^\alpha v_{cb}^\beta v_{ba}^\gamma}{\epsilon_{ac}-\hbar\Omega}\right)\delta(\epsilon_{ab
    }+\hbar\Omega)
    \nonumber\\
    &+
    \frac{v_{ca}^\alpha v_{bc}^\beta v_{ab}^\gamma }{\epsilon_{ac}+\hbar\Omega}\delta(\epsilon_{ab}+2\hbar\Omega)
    \Bigg],
\end{align}
where we defined $[\dd \vb*{k}]\equiv \dd \vb*{k}/(2\pi)^d$ with the spatial dimension $d$, $\epsilon_{ab}=\epsilon_a-\epsilon_b$ and $f_{ab}=f(\epsilon_a)-f(\epsilon_b)$ represent the difference of band energies and the Fermi distribution functions, respectively, and
$v_{ab}^\alpha=\bra{a}\pdv{H_0}{k_\alpha}\ket{b}$ represents the matrix element of the velocity operator.
Here we defined $
\sigma_{\rm BCL}^{\mu\alpha\beta \gamma}
    \equiv \sum_{\{\alpha,\beta,\gamma\}}
    \sigma^{\mu \alpha\beta \gamma}(0;\Omega,\Omega,-2\Omega)
$ by adding up the nonlinear conductivities $\sigma^{\mu\alpha\beta\gamma}$ in all possible permutations of $\{\alpha,\beta,\gamma\}$, e.g. $\sigma_{\rm BCL}^{xxyy}=\sigma^{xxyy}+\sigma^{xyxy}+\sigma^{xyyx}$.
This simplifies the expression due to some cancellations between different tensor elements $\sigma^{\mu \alpha\beta \gamma}$.
We note that the above expression is only valid when the photocurrent is generated by the $C_3$-symmetric BCL driving and $E^{(\Omega)}_\alpha E^{(\Omega)}_\beta E^{(-2\Omega)}_\gamma$ remains the same value regardless of the permutations of the spatial indices $\{\alpha,\beta,\gamma\}$ 
because $\vb*{E}^{(\Omega)}$ and $\vb*{E}^{(-2\Omega)}$ are proportional to each other.

In the two band limit ($n=1,2$), the expressions for $J^x(\Omega)$ and $J^y(\Omega)$ reduce to
\begin{align}\label{eq:twobandlimit}
   J_{\rm BCL}^\alpha(\Omega)&=\frac{2\pi e^4 A_0^3}{3\hbar^5 \Omega\gamma_0} C
   \int [\dd \vb*{k}] f_{12}\Delta v^\alpha
   \nonumber\\
    &\times\biggl[(|v_{12}^x|^2-|v_{12}^y|^2) \Delta v^\alpha + 2 s\Re[v_{12}^xv_{21}^y]\Delta v^\beta \biggr]
   \nonumber\\
    &\times \Bigl(-
    \frac{1}{2}\delta(\epsilon_{12}+\hbar\Omega)+\delta(\epsilon_{12}+2\hbar\Omega)
    \Bigr),
\end{align}
with $(\alpha, \beta, C, s)=(x,y, \cos \theta, -1)$ for $J^x$ and
$(\alpha, \beta, C, s)=(y, x, \sin \theta, 1)$ for $J^y$.
Here, $\Delta v^\alpha=(v_{11}^\alpha-v_{22}^\alpha)$ is the group velocity difference for the two bands. 
In particular, the above expression clearly indicates that one can control the direction of the photocurrent by tuning the phase difference $\theta$ since $J_{\rm BCL}^x\propto \cos\theta$ and $J_{\rm BCL}^y\propto \sin\theta$.



\begin{figure}[t]
\begin{center}
\includegraphics[width=8cm]{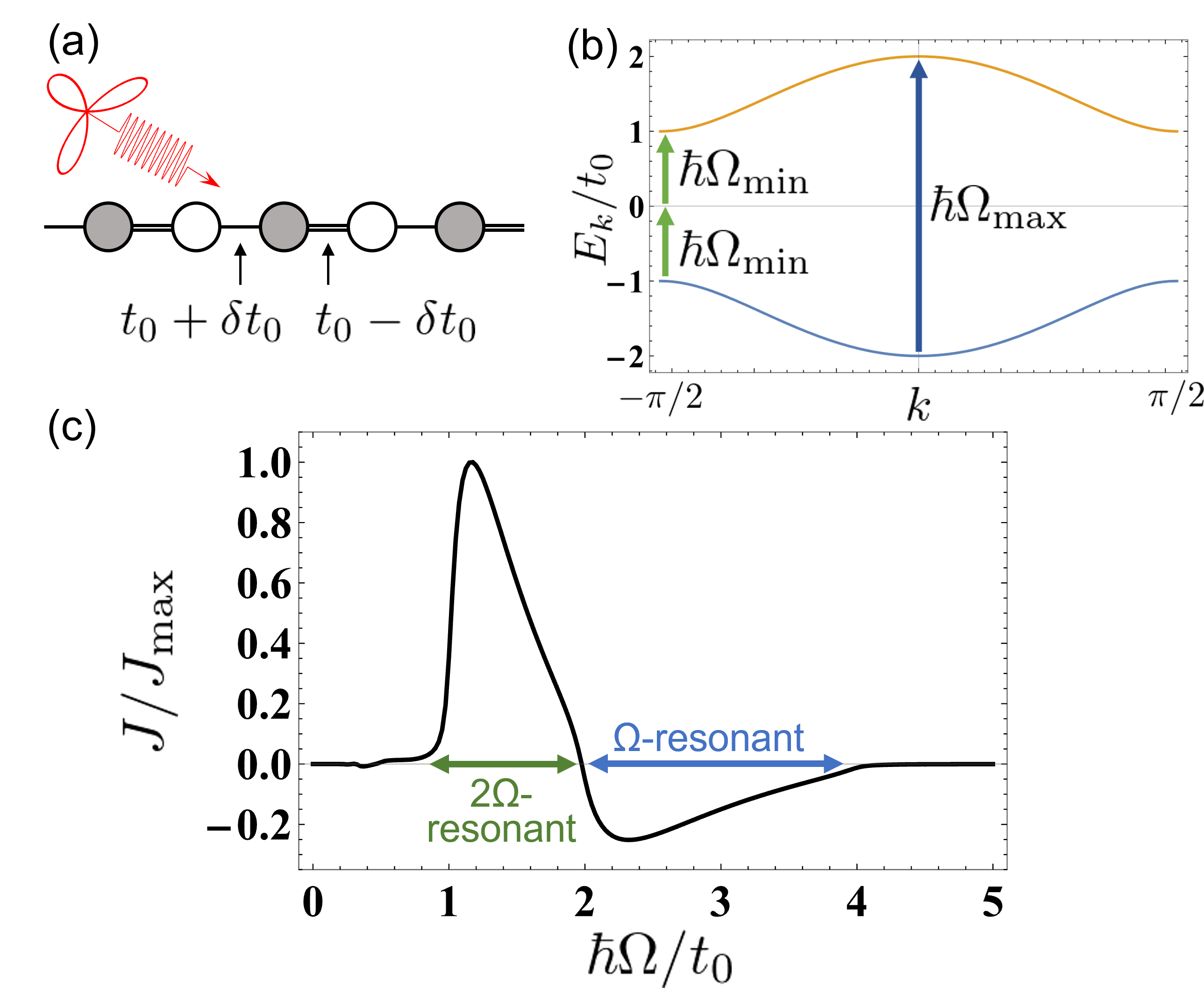}
\caption{BCL-induced photocurrent in 1D SSH model.
(a) A schematic picture of SSH model driven by 3-fold BCL.
(b) The band dispersion of the SSH model. The minimum/maximum input frequency $\Omega_{\rm min/max}$ is depicted with the green/blue arrow. 
(c) Photocurrent dependence on the input frequency. (We set $\delta t_0=0.5t_0$). 
}
\label{fig:SSH1}
\end{center}
\end{figure}

\textit{Applications to 1D SSH model and 3D Dirac/Weyl semimetals. ---}
To demonstrate the BCL-induced photocurrent, we first apply the above formula to the 1D Su-Schrieffer-Heeger (SSH) model, which we adopt as a simple 1D model for systems with inversion symmetry and is described by the Hamiltonian (we set lattice constant to be $a = 2$),
\begin{align}
    H(k)=2t_0 \cos k \sigma_x -2\delta t_0 \sin k\sigma_y,
\end{align}
where $t_0\pm\delta t_0$ is the amplitude of the nearest neighbor hopping with bond alternation and $\sigma_i$ $(i=x,y,z)$ are Pauli matrices [Fig.~\ref{fig:SSH1}(a)].
The energy dispersion is given by 
$
E(k) =\pm \sqrt{4t_0^2 \cos^2 k + 4 \delta t_0^2 \sin^2 k}
$
as plotted in Fig.~\ref{fig:SSH1}(b). 
We note that $H(k)$ has an inversion symmetry, i.e., $\sigma_x H(k) \sigma_x=H(-k)$ which prohibits the conventional second order contribution to the photocurrent and makes the BCL-induced photocurrent a dominant contribution.

Using the formula for two band systems in Eq.~(\ref{eq:twobandlimit}), we obtain the photocurrent for the SSH model under the BCL as
\begin{align}\label{eq:1Dcurrent}
    J_{\rm BCL}(\Omega)
    &= \frac{2\pi e^4 A_0^3}{3\hbar^5 \Omega\gamma_0} \cos\theta \int [\dd k] |v_{12}|^2 (v_{11}-v_{22})^2\\ \nonumber
     &\times \biggl( -\frac{1}{2}\delta(\epsilon_{12}+\hbar\Omega) + \delta(\epsilon_{12}+2\hbar\Omega)
    \biggr).
\end{align}
This shows that the photocurrent induced by BCL driving involves two interband resonance terms:
the $\Omega$-resonant term
$\propto \delta(\epsilon_{12}+\hbar\Omega)$ and 
the $2\Omega$-resonant term
$\propto\delta(\epsilon_{12}+2\hbar\Omega)$ with a sign change.
Figure~\ref{fig:SSH1}(c) shows the photocurrent induced by BCL driving in the 1D SSH model when the Fermi energy lies within the energy gap.
For $|\delta t_0| < |t_0|$, the photo-excitation by BCL driving occurs in the region $2|\delta t_0|\leq \hbar \Omega \leq 2|t_0|$ due to the $2\Omega$-resonant term, and $4|\delta t_0|\leq \hbar \Omega \leq 4|t_0|$ due to the $\Omega$-resonant term. 
The directions of the photocurrent are opposite for the two regions reflecting the relative sign in the two delta functions in Eq.~(\ref{eq:1Dcurrent}). 
The minimum and maximum frequencies for photocurrent generation are given by 
$\hbar \Omega_\textrm{min}=2|\delta t_0|$ and 
$\hbar \Omega_\textrm{min}=4|t_0|$.
Also, we find that the intensity of the first peak is four times larger than the second one due to the $1/\Omega$ term in Eq.~(\ref{eq:1Dcurrent}).


\begin{figure}[t]
\begin{center}
\includegraphics[width=8cm]{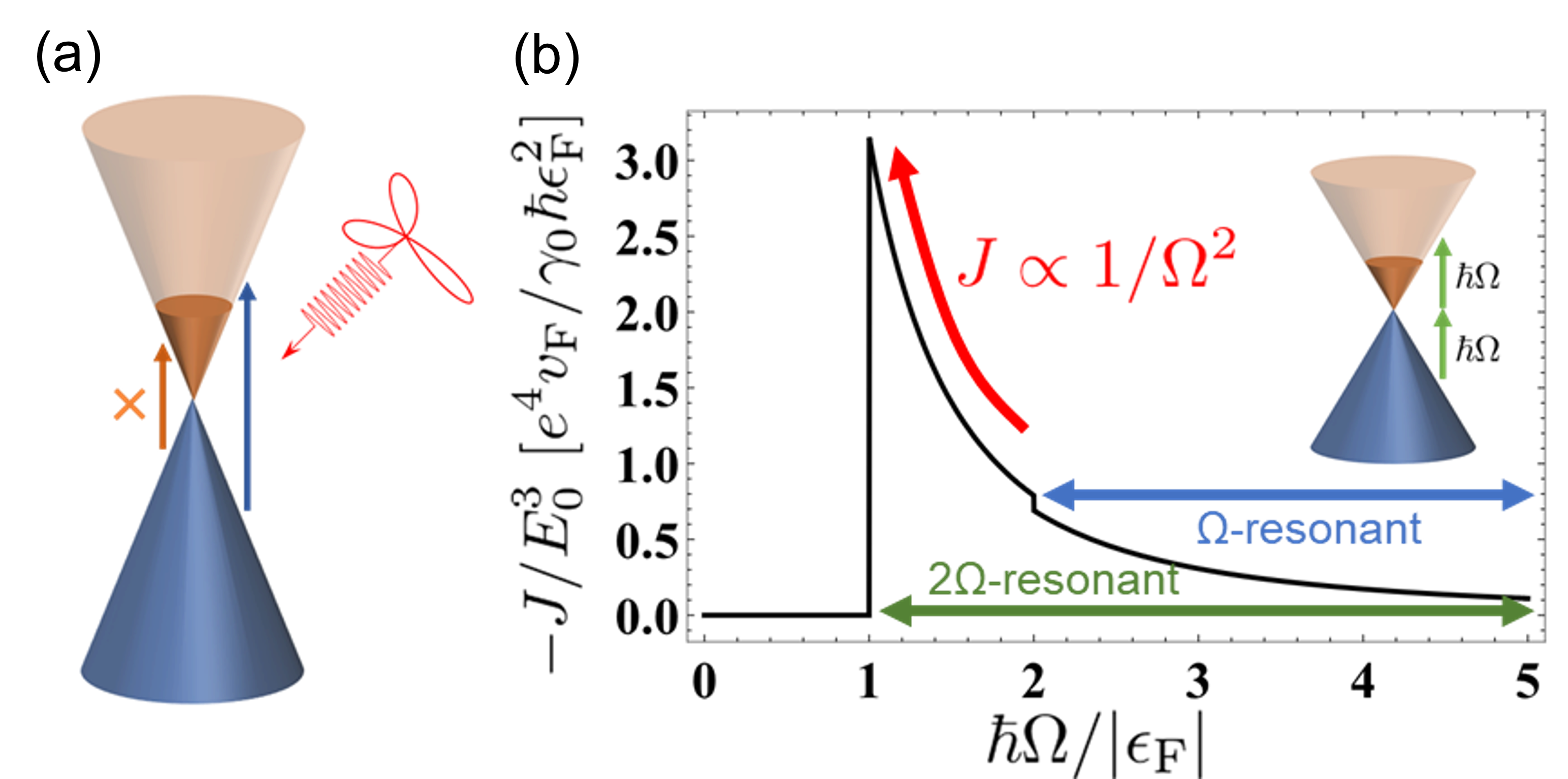}
\caption{BCL-induced photocurrent in Dirac/Weyl semimetals.
(a) A schematic picture of a single Dirac cone driven by 3-fold BCL. 
Pauli blocking prohibits the photo-excitation with the $\Omega$ ($2\Omega$) resonant process for frequencies $\hbar\Omega<2|\epsilon_{\rm F}|$ ($2\hbar\Omega<2|\epsilon_{\rm F}|$) with the chemical potential $\epsilon_{\rm F}$. 
The orange line indicates Pauli blocking for the $\Omega$-resonant process.
(b) Photocurrent dependence on the frequency. We set $\eta=0.8$.
The green (blue) arrow represents the frequency range where the $\Omega$ ($2\Omega$) resonant process is allowed.}
\label{fig:weyl}
\end{center}
\end{figure}

Next we study the BCL-induced photocurrent in 3D Dirac/Weyl semimetals which host gapless linear dispersions.
For simplicity, we consider a model of a single Weyl fermion with anisotropic velocity along the $z$ direction, 
\begin{align}
    H&= \hbar v_{\rm F} (k_x \sigma_x + k_y \sigma_y + \eta k_z \sigma_z).
\end{align}
Here $v_{\rm F}$ is a Fermi velocity and $\eta$ denotes the anisotropy along the $z$ direction.
This model has two-fold rotational symmetry along the $y$ direction $C_2^y$.
Thus, irradiation of $C_3$-symmetric BCL  perpendicular to the (010) surface induces the photocurrent in the $x$-$z$ plane.
From the two-band formula (\ref{eq:twobandlimit}), we obtain
\begin{align}
    \begin{pmatrix}
    J^x_{\rm BCL}(\Omega) \\
    J^z_{\rm BCL}(\Omega) 
    \end{pmatrix}
    &=\frac{4 e^4 |v_{\rm F}|E_0^3}{45\pi \gamma_0 \hbar^2 \Omega^2} \frac{1-\eta^2}{|\eta|} \nonumber\\
    &\times
    \qty[-\Theta(\hbar \Omega- |\epsilon_{\rm F}|)+\frac{1}{8}\Theta(\hbar \Omega- 2|\epsilon_{\rm F}|)]
    \begin{pmatrix}
    \cos\theta \\
    \eta^2\sin\theta
    \end{pmatrix},
\end{align}
where $\Theta(x)$ is the step function.
Note that when the anisotropy is absent with $\eta^2=1$, the system has continuous rotational symmetry.
In this case, the overall driven system still possesses the $C_3^y$ symmetry of the BCL and the photocurrent vanishes.
For a finite chemical potential, the photo-absorption occurs only in the range $\hbar \Omega >|\epsilon_{\rm F}|$ due to the Pauli blocking.
The photocurrent behaves as $J\propto 1/\Omega^2$ with the input light frequency $\Omega$.  
Since Weyl/Dirac semimetals are gapless, this suggests an enhancement of photocurrent by applying the BCL wave in the low frequency region, as shown in Fig.~\ref{fig:weyl}. 
Also the direction of the photocurrent can be controlled with the phase $\theta$ of BCL,
where $(J^x, J^z)$ draws an ellipse by changing the relative phase $\theta$ with the ratio of the minor and major axes being $\eta^2$.
Since this photocurrent is independent of the sign of the Weyl charge (i.e.  sgn($\eta$)),  the photocurrent simply doubles in the Dirac semimetals.

\textit{Nonperturbative effects with the Floquet approach. ---}
Next let us study nonperturbative effects 
for the BCL-induced photocurrent by using an approach based on the Floquet theory.
The Floquet theory describes periodically driven systems by an effective band theory with a Floquet Hamiltonian.
The Floquet approach for nonlinear optical responses treats the optical resonance as a nonequilibrium steady state realized at an anticrossing of the Floquet bands, and is able to incorporate nonperturbative effects with respect to the electric field that cannot be captured by diagrammatic approach \cite{Morimoto-Nagaosa16}.

We study the nonperturbative effect focusing on the SSH model driven by the $C_3$-symmetric BCL. 
The nonequilibrium steady state given by the time-dependent Hamiltonian $H(t)=H_0(k+eA(t)/\hbar)$ with the vector potential of the $C_3$-symmetric BCL 
can be described by the following block of the Floquet Hamiltonian
\begin{align}
    \mathcal{H}_{\rm F}
    =\mqty(
    H_0+2\hbar\Omega & A'v & A' e^{-i\theta}v\\
    A' v & H_0+\hbar\Omega & A' v\\
    A'  e^{i\theta}v  & A' v& H_0
    ),
\end{align}
with $A'=eA_0/2\hbar$.
Here, we consider 
Floquet bands made of photon-dressed  states with 0, 1, and 2 photons,
since the $C_3$-symmetric BCL includes  $\Omega$ and $2\Omega$ frequency components.
Photocurrent in the driven system can be obtained by computing $ie\mathrm{Tr} (v_F G^<)/\hbar$ as detailed in Appendix.
The lesser Green's function $G^<$ encodes information of occupation of Floquet bands,
where the nonequilibrium distribution function of electrons is stabilized by attaching an effective particle bath to the system.
The current operator under the driving $v_{\rm F}$ is defined as $v_{\rm F}=\partial \mathcal{H}_{\rm F}/ \partial k$ when 
$\mathcal{H}_{\rm F}$ is represented with $k$-independent basis.

The photocurrent in the driven systems includes several contributions depending on which Floquet bands are involved in photoexcitation.
One typical contribution for the photocurrent from the $2\Omega$-resonant process is obtained by focusing on three Floquet bands that 
consist of the valence band with one photon, the valence band with two photons, and the conduction band, which reads
\begin{align}\label{eq:Floquet_1D}
    J&=\frac{2\pi eA'^3}{\hbar^2\Omega} \cos\theta \int [\dd k] \frac{|v_{12}|^2 v_{11}(v_{11}-v_{22})}{\gamma'}\delta(\epsilon_{12}+2\hbar\Omega), 
\end{align}
\begin{align}\label{eq: gamma'}
    \gamma' 
    &=\sqrt{\qty|A'e^{-i\theta}v_{12}+\frac{A'^2}{\hbar\Omega}v_{11}v_{12}|^2+\gamma_0^2},
\end{align}
where 
$\gamma'$ is the effective relaxation rate in the presence of the BCL with $\gamma_0$ being the relaxation rate from the fermionic heat bath 
(For details of the derivation, see Appendix).
The factor $A'^3/\gamma'$ gives a nonperturbative effect with respect to $E$.
This leads to a saturation of photocurrent for large intensity with a crossover from $J\propto A_0^3/\gamma_0$ to $J\propto A_0^n$.
The exponent $n$ in the high intensity region is expected to depend on which term in $|\cdots|^2$ becomes dominant in Eq.~\eqref{eq: gamma'}.
While we have discussed the contribution from a specific three-band model in Eq.~(\ref{eq:Floquet_1D}), contributions from other combinations of Floquet bands also show similar crossover behaviors. 
In addition, by adding up all the contributions and focusing on the leading order in $E$, we can reproduce the formula for the BCL-induced photocurrent that was derived from the diagrammatic approach
(For details of the full Floquet formulation, see Appendix).

\begin{figure}[t]
\begin{center}
\includegraphics[width=9cm]{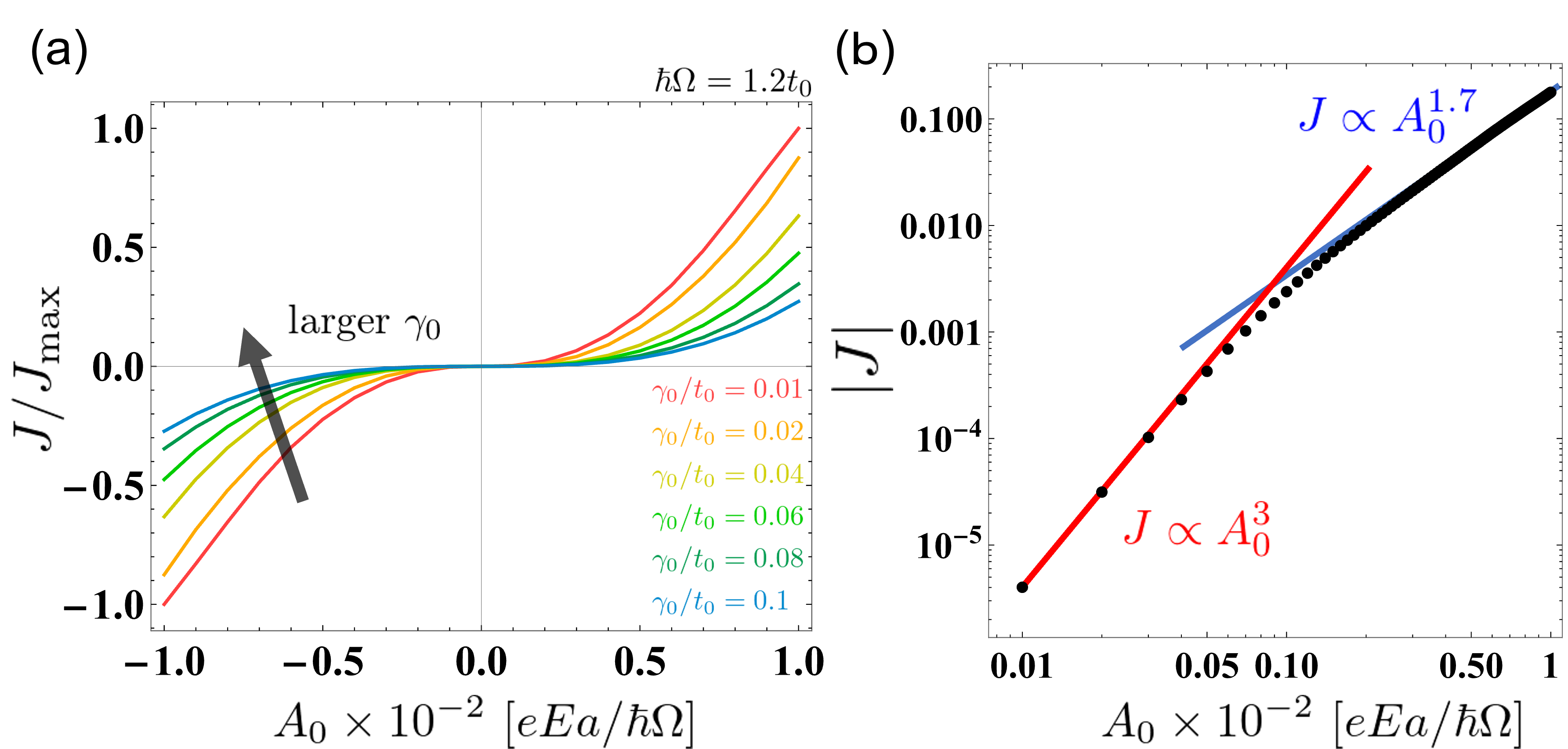}
\caption{Nonperturbative effect on the BCL-induced photocurrent.
(a) The amplitude dependence of the photocurrent with several relaxation rate values $\gamma_0$. 
(b) The amplitude dependence of the photocurrent at $\gamma_0/t_0=0.02$ with fitting curves of $J\propto A_0^3$ (red line) and $J\propto A_0^{1.7}$ (blue line). We set $\hbar\Omega/t_0=1.2$.
}
\label{fig:SSH2}
\end{center}
\end{figure}

Application of the above Floquet formulation to the 1D SSH model shows a nonperturbative correction to the photocurrent (Fig.~\ref{fig:SSH2}).
Figure~\ref{fig:SSH2}(a) shows the amplitude dependence of the photocurrent with several relaxation rate values $\gamma_0$. At small $\gamma_0$, the crossover of photocurrent from $J\propto A_0^3/\gamma_0$ to $J\propto A_0^{1.7}$ occurs as shown in Fig.~\ref{fig:SSH2}(b).
The exponent 1.7 in the high intensity region falls in between 1 and 2 as expected from Eq.~\eqref{eq: gamma'}.
We note that the present Floquet approach can be also applied to 2D/3D systems, where the BCL-induced photocurrent is obtained in a similar way and shows a similar saturation effect.

\textit{Discussions. ---}
The BCL-induced photocurrent is a third order response with respect to the electric field, $J\propto A^3$, as seen in Eq.~\eqref{eq:twobandlimit}.
While the presence of inversion symmetry forbids the second order responses such as the BPVE with monochromatic light, the third order responses are not forbidden in centrosymmetric systems generally. 
In this regard, the appearance of the BCL-induced photocurrent is not contradicting the presence of inversion symmetry in the unperturbed system.
Our finding is that application of the BCL introduces a polarity to the centrosymmetric system and the phase of the two frequency lights can control the direction of the polarity, and hence, the photocurrent.
Such controllability of BCL paves a novel venue for searching optoelectronic functionality.

Since Dirac/Weyl semimetals show a diverging photocurrent at low frequencies, those topological semimetals are a promising platform to observe the BCL-induced photocurrent. 
In particular, a tetragonal 3D Dirac semimetal Cd$_3$As$_2$~\cite{Neupane2014CdAs,Liu2014CdAs} and a hexagonal 3D Dirac semimetal Na$_3$Bi~\cite{Liu2014NaBi} will be good candidate materials.
Both materials host 3D Dirac fermions with anisotropic Fermi velocities along the $z$ direction and the anisotropies are reported to be $\eta\sim 0.25$. 
In these cases, the magnitude of the BCL-induced photocurrent density is estimated as $J\sim 10^6\, \si{A/m^2}$ with realistic parameters $\hbar \Omega=0.1 \, \si{eV}$, $E_0=1 \,$kV/cm, $v_{\rm F}=10^6\,$m/s and $\tau=\hbar/\gamma_0=1\,$ps. 
Considering a sample with the width $L=100\,$\si{\micro \meter} and penetration depth $\delta=1\,$\si{\micro \meter}, the magnitude of the photocurrent is estimated as $I=J L \delta\sim 100\,$\si{\micro \ampere}, which gives a large photocurrent in the mid-infrared region.
Such enhancement of $J$ originates from the gapless nature of Dirac semimetals.
Finally, the field strength exhibiting the crossover due to the nonperturbative effect can be estimated as follows. 
From Fig.~\ref{fig:SSH2}(b), we can estimate the value at which the photocurrent begins to deviate from $J\propto A_0^3$ as $A_0 \sim 0.01 ~[e E_0 a/ \hbar \Omega]$
corresponding to $E_0=0.4\,$MV/cm with a typical lattice constant $a=3\,$\AA~ and a photon energy $\hbar \Omega=1.2\,$eV.

\acknowledgements
We thank Takashi Oka, Masamitsu Hayashi and Ryo Shimano for fruitful discussions.
This work was supported by 
JSPS KAKENHI Grant 20K14407 (S.K.) and 23H01119 (T.M.), 
JST CREST (Grant No. JPMJCR19T3) (S.K., T.M.),
and JST PRESTO (Grant No. JPMJPR19L9) (T.M.).

\bibliography{reference.bib}

\clearpage

\appendix
\begin{widetext}
\section{Appendix A: Diagrammatic approach for BCL-induced photocurrent}

In this section, we show the details of the diagrammatic calculation of the photocurrent induced by BCL driving.
The electron driven by the BCL undergoes a time evolution with the time-dependent Hamiltonian 
\begin{align}
    H(t)=H_0(\vb*{k}+e\vb*{A}(t)/\hbar),
\end{align}
where $H_0(\vb*{k})$ is an unperturbed Hamiltonian.
Here we write the charge of an electron as $-e$ with $e>0$.
We consider the $C_3$-symmetric BCL represented as a complex field for the vector potential,
\begin{align}
    \vb*{A}(t)=\Re[\vb*{A}(e^{i\Omega t}+e^{-i(2\Omega t-\theta)})]
\end{align}
with the complex gauge amplitude $\vb*{A}=A_0(1,-i)$.
The complex electric field is given by
\begin{align}
        \vb*{E}(t)&=-\pdv{\vb*{A}(t)}{t}=\Re [-i\Omega \vb*{A}(e^{i\Omega t}-2e^{-i(2\Omega t-\theta)})]\nonumber \\
        &\equiv \Re [\vb*{E}^{(\Omega)} e^{i\Omega t}+\vb*{E}^{(-2\Omega)} e^{-2i\Omega t}],\\
    \vb*{E}^{(\Omega)}&=-\Omega A_0 \mqty( i \\ 1 ),\quad
    \vb*{E}^{(-2\Omega)}=2e^{i\theta}\Omega A_0 \mqty( i \\ 1 ).
\end{align}
BCL driving can break time-reversal symmetry and spatial symmetries including inversion. 
Thus, the BCL driving can induce photocurrent even in systems with spatial inversion or rotational symmetry.
The BCL-induced photocurrent
\begin{align}
    J^{\mu\alpha\beta\gamma}_{\rm BCL}(\Omega)=\Re \left[
    \sigma^{\mu\alpha\beta\gamma}_{\rm BCL}
     E^{(\Omega)}_\alpha E^{(\Omega)}_\beta E^{(-2\Omega)}_\gamma    \right]
\end{align}
is a third-order response with respect to the electric field at the lowest order since $\Omega+\Omega-2\Omega=0$.
The dominant contribution to the photocurrent is the so called injection current contribution which is proportional to the relaxation time $\tau\sim 1/\gamma_0$.
In the diagrammatic approach, such contribution is included in the box diagram shown in Fig.~\ref{fig:diagram}.
Four black lines stand for the electron propagator
\begin{align}
    G_a (i\omega)=\frac{1}{i\omega - \epsilon_a},
\end{align}
where $\epsilon_a$ is the $a$-th original band energy of the system and $i\omega$ is the Matsubara frequency.
Three black vertices represent the one-photon inputs and we introduce the relaxation rate $\gamma_0$ in the photon energies. The other vertex represents the one-photon output with zero total frequency.

From the Feynman rules, the contribution of this diagram can be written down as
\begin{align} \label{eq: sigma appendix}
    \sigma^{\mu\alpha\beta\gamma}(0;\Omega,\Omega,-2\Omega)&=
    \frac{-ie^4}{2\hbar^4\Omega^3}\mathcal{S}\sum_{abcd}\int[\dd\vb*{k}]v_{ba}^\alpha v_{cb}^\beta v_{dc}^\gamma v_{ad}^\mu I_4(\hbar\Omega+i\gamma_0,\hbar\Omega+i\gamma_0,-2\hbar\Omega+i\gamma_0),
\end{align}
where $v_{ba}^\alpha=\bra{b}\pdv{H_0}{k_\alpha}\ket{a}$ represents the matrix component of the velocity operator and $\mathcal{S}$ denotes the summation for all possible permutations of input photons $(\alpha,\Omega),(\beta,\Omega),(\gamma,-2\Omega)$.
Also we defined a shorthand notation for the $\bm k$ integral as $[\dd\vb*{k}]\equiv \dd\vb*{k}/(2\pi)^d$ with the spatial dimension $d$.
The frequency integral in the box diagram, which we call $I_4$, is performed for imaginary-time Green's functions as \cite{Parker19}
\begin{align}
\begin{split}
    I_4(i\Omega_1,i\Omega_2,i\Omega_3)&=\int \frac{\dd\omega}{2\pi}
    G_a(i\omega)G_b(i\omega+i\Omega_1)G_c(i\omega+i\Omega_1+i\Omega_2)G_d(i\omega+i\Omega_1+i\Omega_2+i\Omega_3)\\
    &=\frac{f(\epsilon_a)}{(\epsilon_{ab}+i\Omega_1)(\epsilon_{ac}+i\Omega_{12})(\epsilon_{ad}+i\Omega_{123})}
        +\frac{f(\epsilon_b)}{(\epsilon_{ba}-i\Omega_1)(\epsilon_{bc}+i\Omega_{2})(\epsilon_{bd}+i\Omega_{23})}\\
        &+\frac{f(\epsilon_c)}{(\epsilon_{ca}-i\Omega_{12})(\epsilon_{cb}-i\Omega_{2})(\epsilon_{cd}+i\Omega_{3})}
        +\frac{f(\epsilon_d)}{(\epsilon_{da}-i\Omega_{123})(\epsilon_{db}-i\Omega_{23})(\epsilon_{dc}-i\Omega_{3})},
\end{split}    
\end{align}
where $\epsilon_{ab}=\epsilon_a-\epsilon_b$, $\Omega_{12}=\Omega_1+\Omega_2$, and $\Omega_{123}=\Omega_1+\Omega_2+\Omega_3$.
$I_4$ in Eq.~\eqref{eq: sigma appendix} is obtained after analytic continuation of Matsubara frequencies,
\begin{align}
    i\Omega_1 &\to \hbar\Omega+i\gamma_0, &
    i\Omega_2 &\to \hbar\Omega+i\gamma_0, &
    i\Omega_3 &\to -2\hbar\Omega+i\gamma_0,
\end{align}
in the above expression.
Note that the Matsubara frequency $i\omega$ is defined such that it has the dimension of the energy.
By taking only the terms proportional to $1/\gamma_0$ in the limit $\gamma_0\to 0$, we can obtain the conductivity of the injection current caused by the band transition from the valence band (denoted as 1) to the conduction band (denoted as 2) as
\begin{align}
    \sigma_{\rm BCL}^{\mu\alpha\beta \gamma}&=\sum_{\{\alpha,\beta,\gamma\}}
    \sigma^{\mu \alpha\beta \gamma}(0;\Omega,\Omega,-2\Omega)\\
    &\simeq \frac{i\pi e^4}{3\hbar^4 \Omega^3\gamma_0}
    \sum_{\substack{\{\alpha,\beta,\gamma\}\\ a,b,c}}\int [\dd \vb*{k}] f_{ab}(v_{aa}^\mu-v_{bb}^\mu)\left[
     \left( \frac{v_{ac}^\alpha v_{cb}^\beta v_{ba}^\gamma}{\epsilon_{ac}+2\hbar\Omega}+\frac{v_{ac}^\alpha v_{cb}^\beta v_{ba}^\gamma}{\epsilon_{ac}-\hbar\Omega}\right)\delta(\epsilon_{ab
    }+\hbar\Omega)+
    \frac{v_{ca}^\alpha v_{bc}^\beta v_{ab}^\gamma }{\epsilon_{ac}+\hbar\Omega}\delta(\epsilon_{ab}+2\hbar\Omega)
    \right],
\end{align}
where $f_{ab}=f(\epsilon_a)-f(\epsilon_b)$ and $a,b,c$ run all of the bands. $\sigma_{\rm BCL}^{\mu\alpha\beta \gamma}$ is defined for clarity by considering cancellations between tensor elements $\sigma^{\mu \alpha\beta \gamma}$ and adding up the nonlinear conductivities $\sigma^{\mu\alpha\beta \gamma}$ in the possible permutations of $\{\alpha,\beta,\gamma\}$, e.g. $\sigma_{\rm BCL}^{xxyy}=\sigma^{xxyy}+\sigma^{xyxy}+\sigma^{xyyx}$.
The photocurrent induced by the $C_3$-symmetric BCL wave can be calculated as
\begin{align}
J^\mu_{\rm BCL}(\Omega)&=\sum_{\{\alpha,\beta,\gamma\}}\Re \left[
    \sigma^{\mu\alpha\beta\gamma}_{\rm BCL}
     E^{(\Omega)}_\alpha E^{(\Omega)}_\beta E^{(-2\Omega)}_\gamma
    \right]\\
    &\equiv \sum_{\{\alpha,\beta,\gamma\}} J^{\mu\alpha\beta\gamma}_{\rm BCL}(\Omega).
\end{align}
We note that this notation is limited to the photocurrent generated by the $C_3$-symmetric BCL driving where $E^{(\Omega)}_\alpha E^{(\Omega)}_\beta E^{(-2\Omega)}_\gamma$ remains the same value regardless of the permutations of the spatial indices $\{\alpha,\beta,\gamma\}$.

\begin{figure}[t]
\begin{center}
\includegraphics[height=4cm]{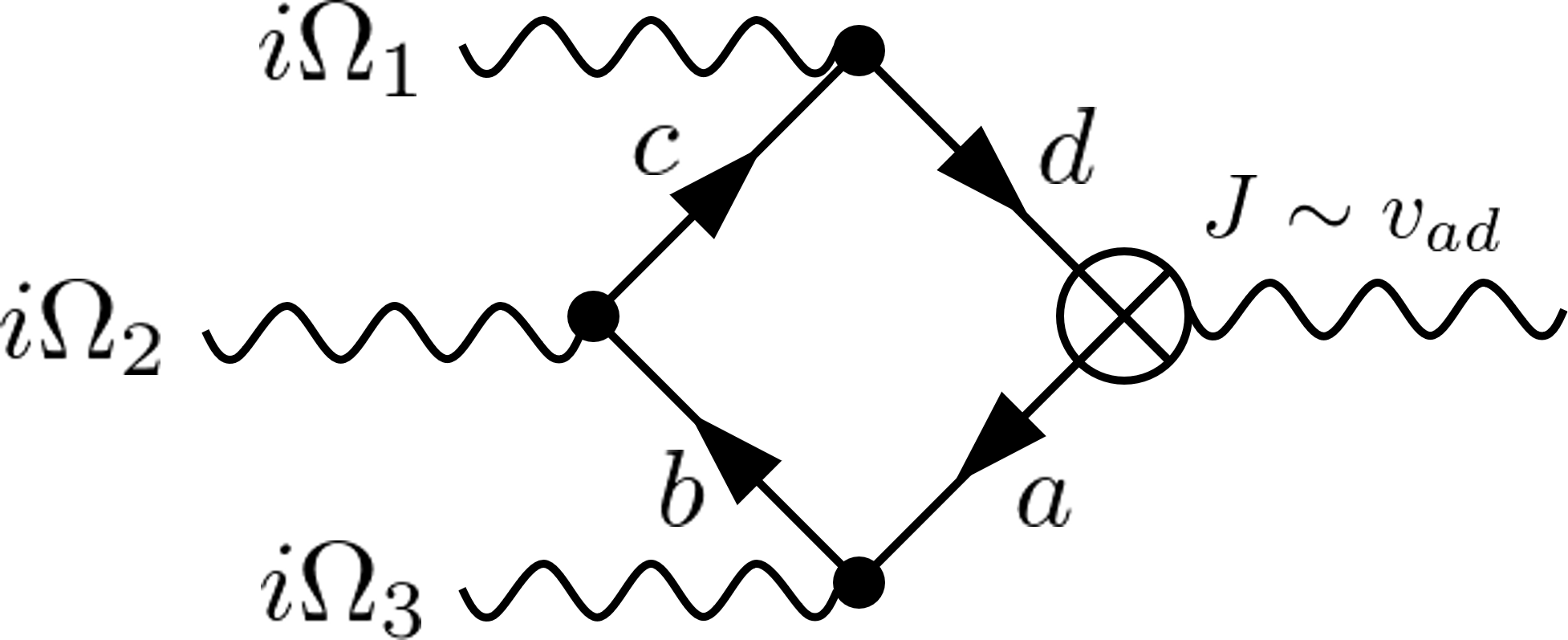}
\caption{The box diagram of the injection current induced by BCL driving.
$\Omega_i=\Omega,-2\Omega$ is the energy of the absorbed photon and satisfies $\sum_i \Omega_i=0$.}
\label{fig:diagram}
\end{center}
\end{figure}

In the two-band limit, the above formulation gives the expressions,
\begin{align} 
   J_{\rm BCL}^x(\Omega)&=\frac{2\pi e^4 A_0^3}{3\hbar^5 \Omega\gamma_0} \cos\theta
   \int [\dd \vb*{k}] f_{12}\Delta v^x
   \biggl[(|v_{12}^x|^2-|v_{12}^y|^2) \Delta v^x - 2\Re[v_{12}^xv_{21}^y]\Delta v^y \biggr]
   \Bigl(-
    \frac{1}{2}\delta(\epsilon_{12}+\hbar\Omega)+\delta(\epsilon_{12}+2\hbar\Omega)
    \Bigr),\\ 
    J_{\rm BCL}^y(\Omega)&=\frac{2\pi e^4 A_0^3}{3\hbar^5 \Omega\gamma_0} \sin\theta
   \int [\dd \vb*{k}] f_{12}\Delta v^y
   \biggl[(|v_{12}^x|^2-|v_{12}^y|^2) \Delta v^y + 2\Re[v_{12}^xv_{21}^y]\Delta v^x \biggr]
   \Bigl(-
    \frac{1}{2}\delta(\epsilon_{12}+\hbar\Omega)+\delta(\epsilon_{12}+2\hbar\Omega)
    \Bigr),
\end{align}
which correspond to
Eq.~(\ref{eq:twobandlimit}) in the main text.
Here, let us further consider a multi-band correction for the BCL-induced photocurrent.
For example, we can write down an explicit formula for one of the current components $J^{xxxx}_{\rm BCL} (\Omega)$ as (for simplicity, the number of bands is set to 3)
\begin{align}\label{eq:Jxxxx}
\begin{split}
    J^{xxxx}_{\rm BCL}(\Omega)&=\frac{2\pi e^4 A_0^3}{3\hbar^5 \Omega\gamma_0} \cos\theta
    \int [\dd \vb*{k}] f_{12} |v_{12}^x|^2(v_{11}^x-v_{22}^x)^2
    \Bigl(-
    \frac{1}{2}\delta(\epsilon_{12}+\hbar\Omega)+\delta(\epsilon_{12}+2\hbar\Omega)
    \Bigr)\\
   &+\frac{2\pi e^4 A_0^3}{3\hbar^4 \gamma_0} 
    \int [\dd \vb*{k}] f_{12} (v_{11}^x-v_{22}^x)\Re[e^{i\theta}v^x_{13}v^x_{32}v^x_{21}]
    \left(
    \frac{1}{\epsilon_{13}+2\hbar\Omega}+\frac{1}{\epsilon_{13}-\hbar\Omega}
    \right)
    \delta(\epsilon_{12}+\hbar\Omega)\\
    &+\frac{2\pi e^4 A_0^3}{3\hbar^4 \gamma_0} \int [\dd \vb*{k}] f_{12} (v_{11}^x-v_{22}^x)
    \Re[e^{i\theta}v^x_{12}v^x_{23}v^x_{31}]
    \frac{1}{\epsilon_{13}+\hbar\Omega}
    \delta(\epsilon_{12}+2\hbar\Omega)\\
    &+[(1,2,3)\leftrightarrow (1,3,2), (2,1,3), (2,3,1),(3,1,2),(3,2,1)].
\end{split}
\end{align}
The first term in Eq.~(\ref{eq:Jxxxx}) represents direct optical transitions from the valence band to the conduction band with the transition rate $|v_{12}^x|^2$. 
The other terms represent optical transitions via the other band (denoted as 3) as the intermediate transition state.
This multi-band correction suggests that a giant photocurrent can be obtained by BCL driving if the band energies are equally separated.
In either term, one can see that the $k$-integral does not vanish even in systems with spatial inversion symmetry, where the conventional photocurrent proportional to $A_0^2$ vanishes.

We note that other BCLs with different combination of $(n_1,n_2)$ can be also utilized to induce photocurrent if they can effectively introduce polarity to the system. 
For the cases $(n_1=1,n_2=n)$, the photocurrent becomes proportional to $E^{n+1}$ in the least order in $E$. 
Thus the photocurrent under BCLs with $n \ge 3$ generally involves higher order optical processes than the third order and tends to require larger strength of the applied electric field.

\section{Appendix B: Floquet-Keldysh approach}
In this section, we present the derivation of the BCL-induced photocurrent by using Floquet-Keldysh formalism.
First, we discuss 1D systems driven by the $C_3$-symmetric BCL.
The time-dependent Hamiltonian is given with the minimal coupling $k\to k+A(t)$ (the convention $\hbar = 1, e = 1$ is used hereafter) as
\begin{align}
    &H(t)=H_0(k+A(t))\simeq H_0(k)+A(t)v,\\
    &A(t)=\Re[A_0(e^{i\Omega t}+ e^{-i(2\Omega t -\theta)})],
\end{align}
where $v=\partial H_0/ \partial k$ is the velocity operator.
Since the Hamiltonian is time-periodic with the period $T=2\pi/ \Omega$, we can apply discrete Fourier transformation and the Floquet picture is given by the Floquet Hamiltonian
\begin{align}
    \mathscr{H}_{\rm F}=\mqty (
\ddots & H_{-1} & H_{-2} & \\
H_1 & H_0-m \Omega & H_{-1} & H_{-2} \\
H_2 & H_1 & H_0-(m+1) \Omega & H_{-1} \\
& H_2 & H_1 & \ddots
),
\end{align}
with Fourier coefficients of the time-dependent Hamiltonian
\begin{align}
    H_m=\frac{1}{T} \int_0^T \mathrm{~d} t e^{i m \Omega t} H(t).
\end{align}
The Floquet approach for nonlinear optical responses treat the optical resonance as a nonequilibrium steady state realized at an anticrossing of the Floquet bands, and is able to incorporate nonperturbative effects with respect to the electric field that cannot be captured by the diagrammatic approach.

Optical responses induced by a conventional monochromatic light can be described by a 2$\times$2 Hamiltonian consisting of only two bands involved in band excitation \cite{Morimoto-Nagaosa16}.
Since the $C_3$-symmetric BCL includes the $2\Omega$ component, the BCL-induced photocurrent  originating from the anticrossing between the valence band and the photon-dressed conduction band can be captured by the truncated Floquet Hamiltonian made of photon-dressed  states with 0, 1, and 2 photons as
\begin{align}
    \mathcal{H}_{\rm F}
    =\mqty(
    H_0+2\Omega & A'v & A' e^{-i\theta}v\\
    A' v & H_0+\Omega & A' v\\
    A'  e^{i\theta}v  & A' v& H_0
    ),
\end{align}
with $A'=A_0/2$.
Photocurrent in the driven systems includes several contributions depending on which Floquet bands are involved in photoexcitation.
Now, to separate them, we further truncate the Floquet Hamiltonian down to 3$\times$3 matrices. For example, we consider one of the Floquet three-band models which consists of the valence band with one photon, the valence band with two photons, and the conduction band as shown in Fig.~\ref{fig:3band}. 
The Hamiltonian of the three-band model is given by
\begin{align}\label{eq:3band}
    H_\text{F}=\mqty(
    \epsilon_1+2\Omega & A'v_{11} & A'e^{-i\theta}v_{12}\\
    A'v_{11} & \epsilon_1+\Omega & A'v_{12}\\
    A' e^{i\theta}v_{21}  & A'v_{21}& \epsilon_2
    ),
\end{align}
where subscripts 1 and 2 represent the valence band and the conduction band respectively as in the previous section.
The photocurrent mainly originates from the anticrossing of the Floquet bands.
Thus, when the interband transition occurs in the $2\Omega$-resonant absorption process $(\epsilon_1+2\Omega\sim \epsilon_2)$, the valence band with one photon can be regarded as a perturbation, denotes as $p$ band, for the anticrossing in the Floquet three band model.
From the Schrieffer-Wolff transformation, the effective Hamiltonian reduced down to a $2\times2$ matrix can be obtained as follows,
\begin{align}
    (H_\text{eff})_{mn}=&(H_{\rm F})_{mn}+\frac{V_{mp}V_{pn}}{2}
    \left(
    \frac{1}{E_m-E_p}+\frac{1}{E_n-E_p}
    \right),\\ \label{eq:Heff}
    H_\text{eff}=&\mqty(
    \epsilon_1+2\Omega + \frac{A'^2}{\Omega}v_{11}^2 &  A'e^{-i\theta}v_{12}+\frac{A^2}{\Omega}v_{11}v_{12}\\
    A' e^{i\theta}v_{21}+\frac{A'^2}{\Omega}v_{11}v_{21}  &  \epsilon_2+\frac{A'^2}{\Omega}|v_{12}|^2
    ),\\
V=&\mqty(
   0 & A'v_{11} & 0\\
    A'v_{11} & \epsilon_1+\Omega & A'v_{12}\\
   0  & A'v_{21}&0
    ),
\end{align}
where $E_n$ represents the quasienergy of the Floquet 3 band model Eq.~(\ref{eq:3band}) and $n,m$ are Floquet band indices excepting the intermediate state $p$.
Here, we assume that the energy denominators are mostly $\Omega$, e.g. $E_1-E_p=E_p-E_2\simeq \Omega$, near the $k$ points where the Floquet bands anticrossing occurs.
\begin{figure}[t]
\begin{center}
\includegraphics[height=6cm]{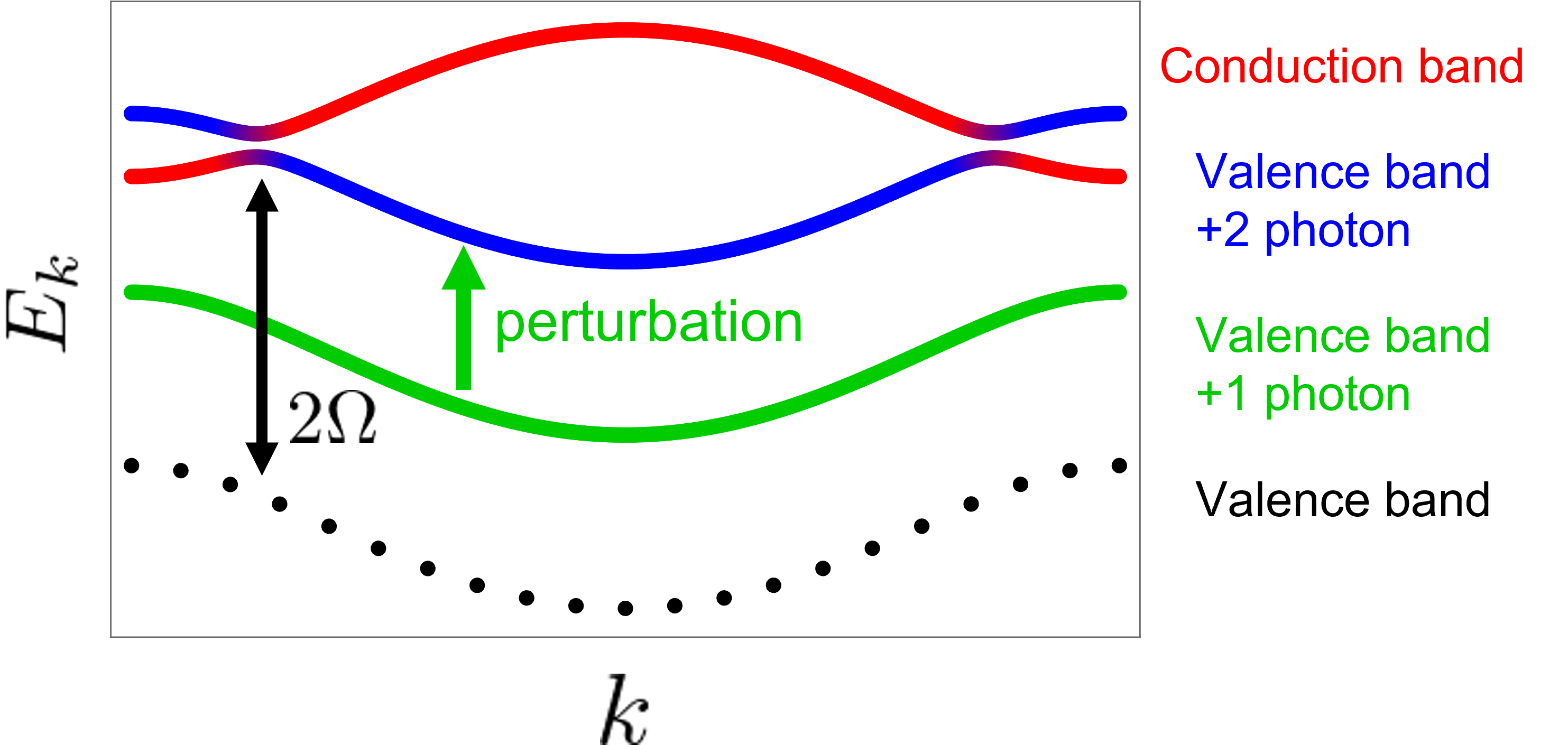}
\caption{Schematic picture of one Floquet three-band model
(depicted by three colored curves).
This three-band model consists of the valence band with one photon, the valence band with two photons, and the conduction band with the band gap $\epsilon_{2}-\epsilon_{1}\sim 2\Omega$.
When photoexcitation occurs in the $2\Omega$-resonant absorption process, the valence band with one photon can be taken as a perturbation for the anticrossing of the Floquet bands.
The black dotted line shows the bare valence band without photon absorption.
We note that this bare valence band is not included in the Floquet three-band model.
}
\label{fig:3band}
\end{center}
\end{figure}
Let us write the Floquet Hamiltonian with two Floquet bands as
$H_{\rm eff} \equiv h_0 + \vb*{h}\vdot \vb*{\sigma}$,
where $\vb*{\sigma}$ are Pauli matrices, and $h_0$ and $\vb*{h}=(h_x,h_y,h_z)$ are coefficients when $H_{\rm eff}$ are expanded with the Pauli matrices.
Optical responses can be obtained from the lesser Green's function $G^<$, which is given for the above two-band model 
as
\begin{align}
    G^<=\frac{(\omega-h_0+i\gamma_0+ \vb*{h}\vdot \vb*{\sigma})\Sigma^<(\omega-h_0-i\gamma_0+ \vb*{h}\vdot \vb*{\sigma})}{[(\omega-h_0+i\gamma_0)^2-h^2][((\omega-h_0-i\gamma_0)^2-h^2]},
\end{align}
where  $h^2=h_x^2+h_y^2+h_z^2$
and $\gamma_0$ is the relaxation rate from the fermionic heat bath. Assuming that each site is coupled to a fermionic heat bath with the coupling constant $2\gamma_0$. The lesser self-energy is given by $\Sigma^<=i\gamma_0(1+\sigma^z)$ when the Fermi energy lies between the valence band and the conduction band.
The photocurrent in the driven system is obtained by directly measuring the expectation value of the velocity operator $v_{\rm F}=\partial H_{\rm eff} /\partial k$ in the nonequilibrium steady state as
\begin{align}
    J = -\int [\dd k] \tr [v_{\rm F} \rho]
\end{align}
with the effective density matrix of the Floquet system
\begin{align}
    \rho = -i \int \frac{\dd \omega}{2\pi} G^<.
\end{align}
Therefore, we can calculate the photocurrent from the following equation as 
\begin{align}
    J&=i \Tr [v_{\rm F}  G^<] 
\equiv i \int \frac{\dd \omega}{2\pi}[\dd k]\mathrm{tr}[v_{\rm F}  G^<].
\end{align}
We note that tr denotes a trace over only band indices. 
For the present 2 band model, this means a simple sum of diagonal components. 
In contrast, for a general Floquet Hamiltonian of a general size, the dc current is defined as
\begin{align}
    J = i \sum_m \Tr [(v_{\rm F})_{nm}  (G^<)_{mn}] 
\end{align}
where $m,n$ represents the Floquet indices,
in order to prevent counting duplicate contributions from different Floquet indices.
One can decompose this into three contributions as $J=\int [\dd k] (j_1+j_2+j_3)$ with \cite{Morimoto-Nagaosa16}
\begin{align}
    j_1&=\frac{\gamma_0 (h_x b_y - h_y b_x)}{h^2 + \gamma_0^2},\\
    j_2&=-\frac{ (h_x b_x + h_y b_y)h_z}{h^2 + \gamma_0^2},\\
    j_3&=-\frac{(h_z^2+\gamma_0^2)b_z}{h^2 + \gamma_0^2}.
\end{align}
Here, we used the velocity operator $v_{\rm F}=b_0+\vb*{b}\vdot\vb*{\sigma}$,
where $b_0$ and $\vb*{b}=(b_x,b_y,b_z)$ are coefficients for the expansion with Pauli matrices. 
Each term $j_i$ corresponds to the resonant shift current, the off-resonant shift current, and the injection current, respectively.
From this formulation, we can obtain the contribution of the BCL-induced injection current from the effective Hamiltonian described by Eq.~(\ref{eq:Heff}) as
\begin{align}
\begin{split}
    J&=-\int [\dd k] \frac{(h_z^2+\gamma_0^2)b_z}{h^2+\gamma_0^2}\\
    &=\int [\dd k] \frac{h_x^2+h_y^2}{h^2+\gamma_0^2}(v_{11}-v_{22})\\
    &=\int [\dd k] \frac{|A'e^{-i\theta}v_{12}+\frac{A'^2}{\Omega}v_{11}v_{12}|^2}{h^2+\gamma_0^2}(v_{11}-v_{22})\\
    &=\frac{2\pi A'^3}{\Omega} \cos\theta \int [\dd k] \frac{|v_{12}|^2 v_{11}(v_{11}-v_{22})}{\gamma'}\delta(\epsilon_{12}+2\Omega).
\end{split}
\end{align}
In the last line, we use the fact that the terms proportional to $A_0^2$ or $A_0^4$ in the numerator vanish in the presence of the inversion symmetry or $C_2$ symmetry.
Also, assuming that the electric field and $\gamma$ are small enough, we can apply the following approximation 
\begin{align}
    \frac{1}{h^2+\gamma_0^2}\sim
    \frac{\pi \delta(h_z)}{\gamma'}
\end{align}
where 
\begin{align}
    \gamma'^2=h_x^2+h_y^2+\gamma_0^2=\left|A'e^{-i\theta}v_{12}+\frac{A'^2}{\Omega}v_{11}v_{12}\right|^2+\gamma_0^2
\end{align}
is the effective relaxation rate. This nonperturbative modification results in a saturation of optical excitations and induces a departure from the $J\propto A_0^3/\gamma$ behavior.
In the high intensity region, the current behaves as $J\propto A_0^n$ with $1\le n\le 2$, depending on which term in $|\cdots|$ becomes dominant by increasing the driving intensity.

So far, we have discussed the contribution from a specific three-band model as shown in Fig.~\ref{fig:3band}. However, other Floquet three bands must be considered. 
In the $2\Omega$-resonant absorption process ($\epsilon_2-\epsilon_1\sim 2\Omega$),
another three-band model can be considered:
\begin{align}
    H_\text{F}=\mqty(
    \epsilon_1+2\Omega & A'v_{12} & A'e^{-i\theta}v_{12}\\
    A'v_{21} & \epsilon_2+\Omega & A'v_{22}\\
    A' e^{i\theta}v_{21}  & A'v_{22}& \epsilon_2
    ),\quad
    H_\text{eff}=
    \mqty(
   \epsilon_1+2\Omega - \frac{A'^2}{\Omega}|v_{12}|^2 &  A'e^{-i\theta}v_{12}-\frac{A'^2}{\Omega}v_{12}v_{22}\\
    A' e^{i\theta}v_{21}-\frac{A'^2}{\Omega}v_{12}v_{22}  &  \epsilon_2-\frac{A'^2}{\Omega}v_{22}^2
    ).
\end{align}

In the $\Omega$-resonant absorption process ($\epsilon_2-\epsilon_1\sim \Omega$),
four three-band models can be taken into account as follows:
\begin{align}
    H_\text{F}&=\mqty(
    \epsilon_1+2\Omega & A'v_{11} & A'e^{-i\theta}v_{12}\\
    A'v_{11} & \epsilon_1+\Omega & A'v_{12}\\
    A' e^{i\theta}v_{21}  & A'v_{21}& \epsilon_2
    ),\quad
    H_\text{eff}=\mqty(
    \epsilon_1+\Omega - \frac{A'^2}{\Omega}v_{11}^2 &  A'v_{12}-\frac{A^2}{\Omega}e^{-i\theta}v_{11}v_{12}\\
    A' v_{21}-\frac{A'^2}{\Omega}e^{i\theta}v_{11}v_{21}  &  \epsilon_2-\frac{A'^2}{\Omega}|v_{12}|^2
    ),\\
    H_\text{F}&=\mqty(
    \epsilon_1+2\Omega & A'v_{12} & A'e^{-i\theta}v_{12}\\
    A'v_{21} & \epsilon_2+\Omega & A'v_{22}\\
    A' e^{i\theta}v_{21}  & A'v_{22}& \epsilon_2
    ),\quad
    H_\text{eff}=
    \mqty(
   \epsilon_1+\Omega + \frac{A'^2}{\Omega}|v_{12}|^2 &  A'v_{12}+\frac{A'^2}{\Omega}e^{-i\theta}v_{12}v_{22}\\
    A' v_{21}+\frac{A'^2}{\Omega}e^{i\theta}v_{22}v_{21}  &  \epsilon_2+\frac{A'^2}{\Omega}v_{22}^2
    ),\\
    H_\text{F}&=\mqty(
    \epsilon_2+2\Omega & A'v_{21} & A'e^{-i\theta}v_{22}\\
    A'v_{12} & \epsilon_1+\Omega & A'v_{12}\\
    A' e^{i\theta}v_{22}  & A'v_{21}& \epsilon_2
    ),\quad
    H_\text{eff}=\mqty(
    \epsilon_1+\Omega - \frac{A'^2}{2\Omega}|v_{12}|^2 &  A'v_{12}-\frac{A^2}{2\Omega}e^{-i\theta}v_{12}v_{22}\\
    A' v_{21}-\frac{A'^2}{2\Omega}e^{i\theta}v_{12}v_{22}  &  \epsilon_2-\frac{A'^2}{2\Omega}v_{22}^2
    ),\\
    H_\text{F}&=\mqty(
    \epsilon_1+2\Omega & A'v_{12} & A'e^{-i\theta}v_{11}\\
    A'v_{21} & \epsilon_2+\Omega & A'v_{21}\\
    A' e^{i\theta}v_{11}  & A'v_{12}& \epsilon_1
    ),\quad
    H_\text{eff}=
    \mqty(
   \epsilon_1+\Omega + \frac{A'^2}{2\Omega}v_{11}^2 &  A'v_{12}+\frac{A'^2}{2\Omega}e^{-i\theta}v_{11}v_{12}\\
    A' v_{21}+\frac{A'^2}{2\Omega}e^{i\theta}v_{21}v_{11}  &  \epsilon_2+\frac{A'^2}{\Omega}|v_{12}|^2
    ).
\end{align}
The effective relaxation rates in calculating the photocurrent from these three-band models are slightly different in the contents of $|\cdots|$ respectively, but remain qualitatively the same and show the same saturation effect.
In the small field region, by adding up all the contributions, we can obtain the same formula for the photocurrent induced by BCL driving as that derived from the diagrammatic approach
\begin{align}\label{eq:1Dphotocurrent}
    J
    \simeq \frac{2\pi A'^3}{\Omega \gamma_0} \cos\theta \int [\dd k] |v_{12}|^2 (v_{11}-v_{22})^2
    \biggl( -\frac{1}{2}\delta(\epsilon_{12}+\Omega) + \delta(\epsilon_{12}+2\Omega)
    \biggr).
\end{align}

Next, we consider 2D systems driven by $C_3$-symmetric BCL, which is given by
\begin{align}
    &H(t)=H_0(\vb*{k}+\vb*{A}(t)),\\
    &A_x(t)+iA_y(t)=A(e^{i\Omega t}+ e^{-i(2\Omega t -i\theta)}).
\end{align}
As well as 1D systems, we can construct three-band models for 2D driven systems.
For example, a 2D Floquet three-band model consisting of the valence band with one photon, the valence band with two photons, and the conduction band is modified as follows:
\begin{align}
    H_\text{F}=\mqty(
    \epsilon_1+2\Omega & A'(v^x_{11}-iv^y_{11}) & A'e^{-i\theta}(v^x_{12}+iv^y_{12})\\
    A'(v^x_{11}+iv^y_{11}) & \epsilon_1+\Omega & AA'(v^x_{12}-iv^y_{12})\\
    A' e^{i\theta}(v^x_{21}-iv^y_{21})  & A'(v^x_{21}+iv^y_{21})& \epsilon_2
    ).
\end{align}
In exactly the same way as above, an effective Hamiltonian for the three-band model can be obtained by performing the Schrieffer-Wolff transformation as
\begin{align}
    H_\text{eff}=\mqty(
    \epsilon_1+2\Omega + \frac{A^2}{\Omega}[(v_{11}^x)^2+(v_{11}^y)^2] &  Ae^{-i\theta}(v^x_{12}+iv^y_{12}) +\frac{A^2}{\Omega} (v^x_{11}-iv^y_{11})(v^x_{12}-iv^y_{12})\\
    Ae^{i\theta}(v^x_{21}-iv^y_{21}) +\frac{A^2}{\Omega} (v^x_{11}+iv^y_{11})(v^x_{21}+iv^y_{21})  &  \epsilon_2+ \frac{A^2}{\Omega}|v_{12}^x-iv_{12}^y|^2
    ).
\end{align}
From the Keldysh Green's function method, the photocurrent can be calculated as
\begin{align}
    J^\alpha = 
    \int [\dd k] \frac{h_x^2+h_y^2}{h^2+\gamma_0^2}\Delta v^\alpha,
\end{align}
with $\Delta v^\alpha=v_{11}^\alpha-v_{22}^\alpha$.
By adding up all possible Floquet three-band models, the formula for the BCL-induced photocurrent derived by the diagrammatic method [Eq.~\eqref{eq:twobandlimit}] can be reproduced.
\clearpage

\end{widetext}
\end{document}